\documentclass[aps,prb,final,twocolumn,letterpaper,superscriptaddress]{revtex4}

\newcommand{\vect}[1]{\mathbf{#1}}
\usepackage{amssymb,amsmath} 
\usepackage{color}
\usepackage[normalem]{ulem}
\usepackage{graphicx}

\begin{document}

\title{Inverse Melting of an Electronic Liquid Crystal}

\author{Shu-Han Lee}
\author{Yen-Chung Lai}
\author{Chao-Hung Du}
\affiliation{Department of Physics, Tamkang University, Tamsui Dist., New Taipei City 25137, Taiwan}
\author{Ying-Jer Kao}
\affiliation{Advanced Center for Theoretical Science and Department of Physics, National Taiwan University, No. 1, Sec. 4, Roosevelt Rd. Taipei, 10607, Taiwan}
\author{Alexander F. Siegenfeld}
\affiliation{Department of Physics, Massachusetts Institute of Technology, 77 Massachusetts Avenue, Cambridge, MA 02139-4307}
\author{Peter D. Hatton}
\affiliation{Department of Physics, Durham University, Durham DH1 3LE, UK}
\author{D. Prabhakaran}
\affiliation{Department of Physics, University of Oxford, Clarendon Laboratory, Parks Road, OX1 3PU, UK}
\author{Yixi Su}
\affiliation{Juelich Centre for Neutron Science JCNS, Forschungszentrum Juelich GmbH, Outstation at MLZ, D-85747, Garching, Germany}
\author{Di-Jing Huang}
\affiliation{National Synchrotron Radiation Research Center, 101 Hsin-Ann Road, Hsinchu 30076, Taiwan}

\date{\today}

\begin{abstract}
Inverse melting refers to the rare thermodynamic phenomenon in which a solid melts into a liquid upon cooling~\cite{Greer:1995}, a transition that can occur only when the ordered (solid) phase has more entropy than the disordered (liquid) phase,   
and that has so far only been observed in a handful of systems~\cite{Rastogi:1999fp,Dobbs:2000kl,Avraham:2001zp,Schupper:2005}.
Here we report the first experimental observation for the inverse melting of an electronic liquid crystalline order~\cite{Kivelson:2003ph,Fradkin:2010qa} in the layered compound La$_{2-x}$Sr$_x$NiO$_4$ (LSNO) at the hole doping concentration $x$ = ${1}/{3}$ .  Using x-ray scattering, we demonstrate that the isotropic charge modulation is driven to nematic order by fluctuating spins and shows an inverse melting transition. Using a phenomenological Landau theory, we show that this inverse melting transition is due to the interlayer coupling between the charge and spin orders. This  discovery points to the importance of the interlayer correlations in the system, and provides a new perspective to study the intricate nature of the electronic liquid crystal phases in strongly correlated electronic systems, including possibly the Cu- and Fe-based high-Tc superconductors~\cite{Fernandes:2014,Lee:2006de}.
\end{abstract}

\maketitle

\section{Introduction}
\vspace{-1em}
Electronic liquid crystal phases arise due to the Coulomb-frustrated separation of electronic domains at the nanoscale~\cite{Kivelson:2003ph,Fradkin:2010qa}. In these  phases, there exist locally Mott insulating regions with magnetic (spin) order,  separated by more metallic regions with higher concentrations of doped holes.  It is well established that there can exist both smectic and striped-liquid phases of in-plane charge and spin ordering in LSNO~\cite{Freeman:2004,Lee:2002qq,Anissimova:2013}.  Here we report x-ray scattering measurements on single-crystal samples of LSNO, which has a tetragonal structure (Fig.~\ref{fig1}a) and is isostructural with the superconducting cuprate LSCO.  

Both LSNO and LSCO are antiferromagnetic (AFM) Mott insulators in the absence of hole doping. While LSCO becomes a high-T$_c$ superconductor for small amounts of hole doping, LSNO remains insulating for doping levels of up to 90\%~\cite{Cava:1991rc}.   In LSNO the doped holes condense, leaving, within each 2D NiO layer, an alternating pattern of AFM domains (spin stripes) separated by charge stripes (Fig.~\ref{fig1}b).  In the reciprocal space of the tetragonal crystal structure ($F4/mmm$), the stripes lead to charge and spin satellite reflections with wavevectors of $\mathbf{Q}_{\rm CO}=(H\pm 2\epsilon\ 0\ L_1)$ and $\mathbf{Q}_{\rm SO}=(H\pm \epsilon\ 0\ L_2)$, where $H$ and $L_2$ are  integers, $L_1$ is odd, and $\epsilon$ is determined by hole concentration with $\epsilon \sim x$.  For La$_{5/3}$Sr$_{1/3}$NiO$_4$ ($x=1/3$), the charge and spin orders are commensurate with the lattice, and satellite reflections from the charge stripes superimpose on those from the spin stripes (Fig.~\ref{fig1}c), a condition that proves essential for the inverse melting of the interlayer charge order. 
 
\section{Measurements}
\begin{figure*}
\includegraphics[width=1.\textwidth]{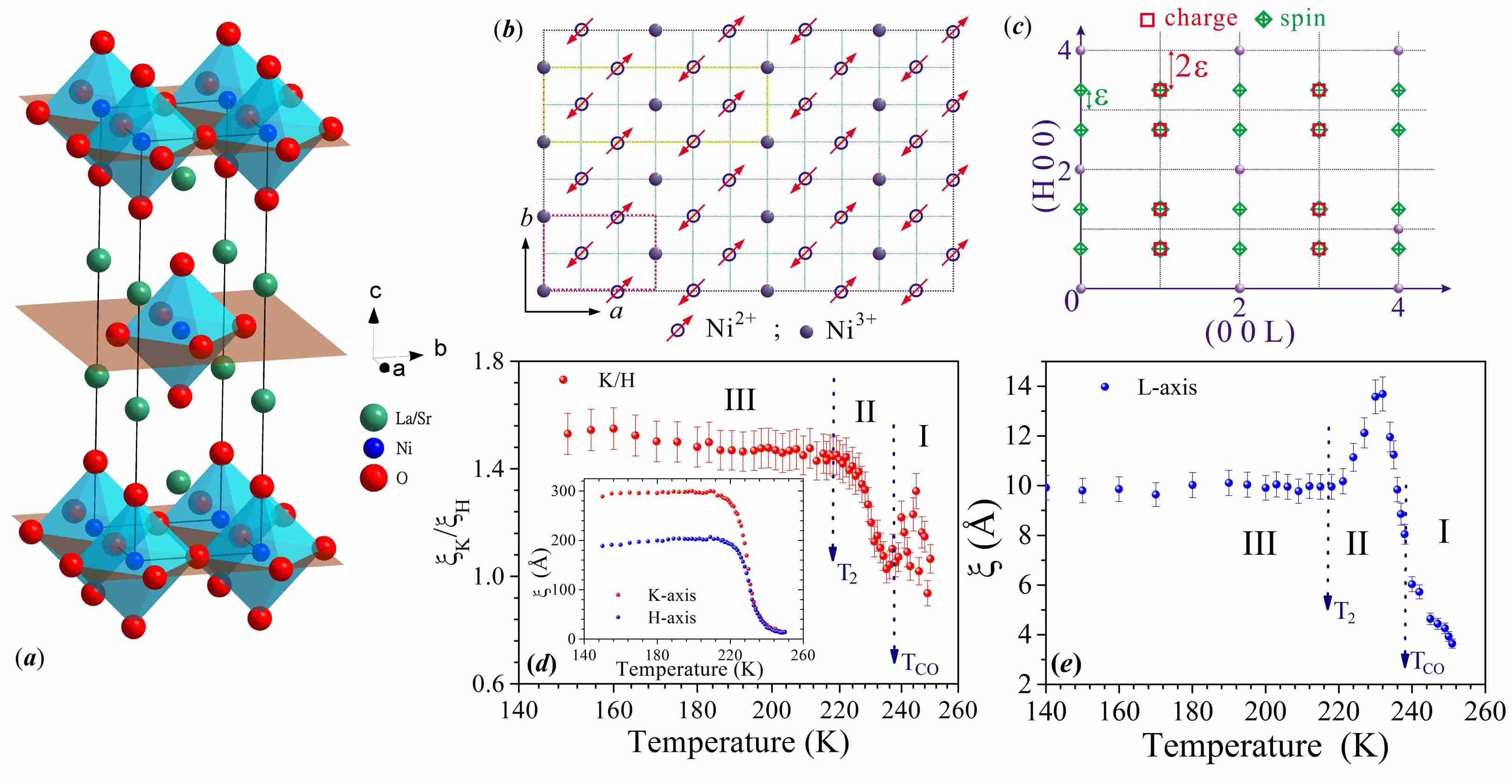}
\caption{Schematic views of crystal structure and charge stripes, and the evolution 
of correlation lengths of charge stripes as a function of temperature. (a) The 
crystal structure of La$_{5/3}$Sr$_{1/3}$NiO$_4$. (b) Schematic view of charge and spin stripes in 
La$_{5/3}$Sr$_{1/3}$NiO$_4$ in NiO$_2$ planes of the tetragonal unit cell. The arrows represent the 
Ni$^{2+}$ ions, and solid circles are the holes. Yellow and red boxes indicate the size of the 
spin and charge modulations. (c) The satellite reflections of charge and spin stripes in 
the $(H~0~L)$ plane of reciprocal space. Since the incommensurability $\epsilon\sim1/3$, charge 
reflection satellites superimpose on spin reflections. (d) Temperature evolution of the 
ratio of the correlation lengths along the $K$ and $H$ directions. The data can be divided 
into 3 regions with two transition temperatures of $T_{CO}$ ($\sim$238 K) and $T_2$ ($\sim$218 K), as 
marked I, II, and III. Below $T_{CO}$, anisotropy in the $H$ and $K$ directions emerges while 
no lattice distortion is present, suggesting electronic nematic- and smectic-liquid 
crystal phases in II and III, respectively. In region I, the system is disordered with no 
anisotropy. The inset shows the evolution of the correlation lengths of the charge 
stripes along the $H$ and $K$ directions as a function of temperature. (e) Evolution of the 
correlation length along the $L$-direction as a function of temperature. As can be seen, 
there is an inverse order-disorder transition at around 230 K.}
\label{fig1}
\end{figure*}

\vspace{-1em}
Figure~\ref{fig1}d shows, as a function of temperature, the ratio of the charge correlation lengths along the $H$ and $K$ directions in the reciprocal space.   
In regime I ($T>238 $K), the system is in an isotropic electronic liquid phase. In regime II (218 K$<T<238$ K), the anisotropy increases  and the ordering can be identified as a nematic phase which breaks the $C_4$ symmetry of the crystal within each layer~\cite{Lee:1997vl,Vojta:2009}. LSNO has been known not to have any lattice distortions at low temperatures~\cite{Abeykoon:2013ai}, so this anisotropic behaviour is a result of an intrinsic charge modulation. In regime III ($T<218$ K), the ratio is nearly constant, consistent with the picture of smectic ordering~\cite{Kivelson:2003ph,Fradkin:2010qa}.

The unusual data comes from the measurements taken along the $L$-direction in the reciprocal space, shown in Fig.~\ref{fig1}e.   
Cooling from high temperatures, the charge correlation starts to build up significantly along the $c$-axis of the crystal at around $T=238$ K, and the charge correlation lengths start to increase from $\xi$= $\sim$6 \AA\  to 14 \AA\   as temperature is cooled down to $T= 230$ K. At this temperature, the interlayer charge correlation spans over  two NiO layers and it seems that a full 3D ordering will eventually develop. However, when the temperature is further decreased, the interlayer charge correlation starts to decrease, rather than increase, and the inverse melting occurs. Finally, the interlayer charge correlation length reaches $\xi$ = $\sim$10  \AA\   below $T$= 218 K, where both the  charge and spin stripes are well established.  Thermal hysteresis behaviour around the transition (see Fig.~\ref{figS1}) matches the expected thermal behaviour of electronic nematic liquids~\cite{Dahmen:2010}. 

\begin{figure}
\includegraphics[width=.5\textwidth]{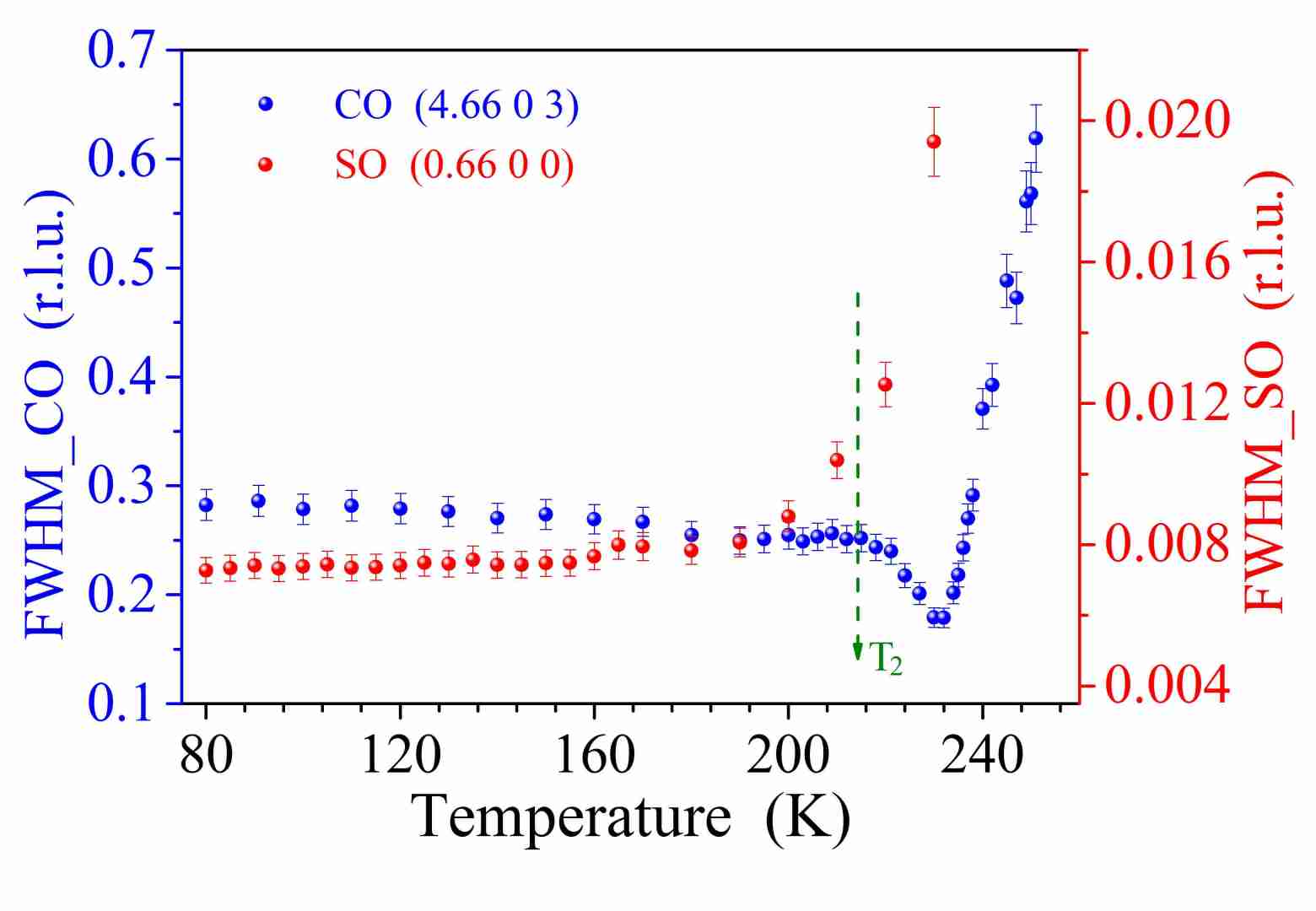}
\caption{Interplay between charge and spin stripes. Evolution of peak width as a 
function of temperature for the charge ordering reflection $(4.66~0~3)$ [blue dots] and 
the spin ordering reflection $(0.66~0~0)$ [red dots]. The charge reflection was measured 
along the $L$-direction using hard x-rays, and the spin reflection was measured along 
the $H$-direction by resonant soft x-ray diffraction. It can be seen that the spin stripe 
order exists up to $\sim$230 K. The onset of the interlayer correlation suppression 
coincides with the onset of the in-plane spin order, indicating a close relationship 
between the two.}
\label{fig2}
\end{figure}

Experiments were also conducted to measure the spin stripes using resonant soft x-ray diffraction.  Figure~\ref{fig2} shows the temperature evolution  of the peak profile of a spin stripe satellite reflection at (0.66\ 0\ 0) measured along the $H$ direction, which serves as a measure of in-plane spin order, and of a charge stripe satellite reflection at (4.66\ 0\ 3) along the $L$ direction, which serves as a measure of interlayer charge correlation.  Although the spin stripe transition occurs at a temperature of around 190 K, the satellite reflection persists up to temperatures as high as 230 K, indicating that between 190 K and 230 K, the material exhibits a spin stripe liquid~\cite{Lee:2002qq}.  The  onset of the interlayer charge order suppression coincides with the appearance of the in-plane spin stripe order, suggesting that the in-plane spin stripe order plays an important role in the inverse melting of the interlayer charge order.

\section{Theoretical Model}
\vspace{-1em}  
In order to model the observed behaviour in LSNO, we construct a Landau theory for the spin and charge stripe orders for a bilayer system with 2D layers.  For simplicity we assume that the spin and charge order in each layer can be described by single complex Fourier coefficients and that the spin order is collinear; thus the order parameters can be written as $\vect{S}_i=|S_i|e^{i(\phi_i+\vect{r}_i\cdot \vect{q}_i^S)}\hat{\vect{m}}_i$ and ${\rho}_i=|\rho_i|e^{i(\theta_i+\vect{r}_i\cdot \vect{q}_i^\rho)}$, where $\vect{q}_i^S$ is measured relative to the in-plane antiferromagnetic ordering vector $\mathbf{Q}=(1,0,0)$ and $i\in\{1,2\}$ denotes the layer index.  We take $2\vect{q}_1^S=2\vect{q}_2^S=
\vect{q}_1^\rho=\vect{q}_2^\rho$, so as to allow coupling between the order parameters within and between layers~\cite{Zachar:1998ov}.

Starting from the most general Landau free energy for a single layer that includes all symmetry allowed terms up to the fourth order and then applying a few simplifications yields~\cite{Zachar:1998ov} 

\begin{widetext}
\begin{equation}
F_i=\frac{1}{2}r_s|S_i|^2+|S_i|^4+
\frac{1}{2}r_\rho|\rho_i|^2+|\rho_i|^4
+\lambda_1|S_i|^2|\rho_i|\cos(2\phi_i-\theta_i).
\label{eq:phase1}
\end{equation}
We take the free energy due to the interlayer coupling to be

\begin{align}
F_c=&\lambda_\rho(\rho_1\rho_2^*+c.c.)+
\lambda_2([(\vect{S_1}\cdot\vect{S_1})\rho_2^*+(\vect{S_2}\cdot\vect{S_2})\rho_1^*]+c.c.)\nonumber\\
=&2\lambda_\rho|\rho_1||\rho_2|\cos(\theta_1-\theta_2)+
2\lambda_2\left[|S_2|^2|\rho_1|\cos(2\phi_2-\theta_1)
+|S_1|^2|\rho_2|\cos(2\phi_1-\theta_2)\right], \label{eq:phase2}
\end{align}
\end{widetext}
where the $\lambda_\rho$ term is due to Coulomb repulsion between layers and the $\lambda_2$ term is due to the fact that the holes are to some extent delocalized between layers.

The total free energy is $F=F_1+F_2+F_c$, but considering that intralayer interactions are far stronger than interlayer ones, the approximate values of $|S_i|$, $|\rho_i|$, and $2\phi_i-\theta_i$ can be determined by examining only the single-layer free energy, resulting in $|S_1|=|S_2|\equiv S$ and $|\rho_1|=|\rho_2|\equiv \rho$, and leaving only the value of $(\theta_1-\theta_2)\equiv\alpha$ to be determined by the interlayer coupling.  Minimizing the intralayer free energies requires $\cos(2\phi_i-\theta_i)=-1$, since  $\lambda_1$ is positive~\cite{Zachar:1998ov}, and so we obtain 
$F_c=2\rho(\lambda_\rho\rho-2\lambda_2 S^2)\cos\alpha$.
A good measure of the strength of the interlayer coupling is the value of $\frac{\partial^2 F_c}{{\partial \alpha}^2}$ at equilibrium, which is $|2\rho(\lambda_\rho\rho-2\lambda_2 S^2)|\equiv C$.  For a large $C$, the phase shift between layers of the charge stripes is pinned, but for small $C$, the phase shift between layers becomes less rigid and allows for more fluctuations, leading to a reduction of  interlayer correlation. It is now clear why the interlayer correlations at first increase and then decrease as the temperature $T$ is lowered: at high temperatures, $\rho=S=0$ so there are no interlayer correlations.  Charge order appears first and so $C$ at first increases, but at lower temperatures spin order also appears thus causing $C$ to then decrease.  At still lower temperatures we expect the Landau theory to no longer accurately model the system.

\begin{figure*}
\includegraphics[width=1.\textwidth]{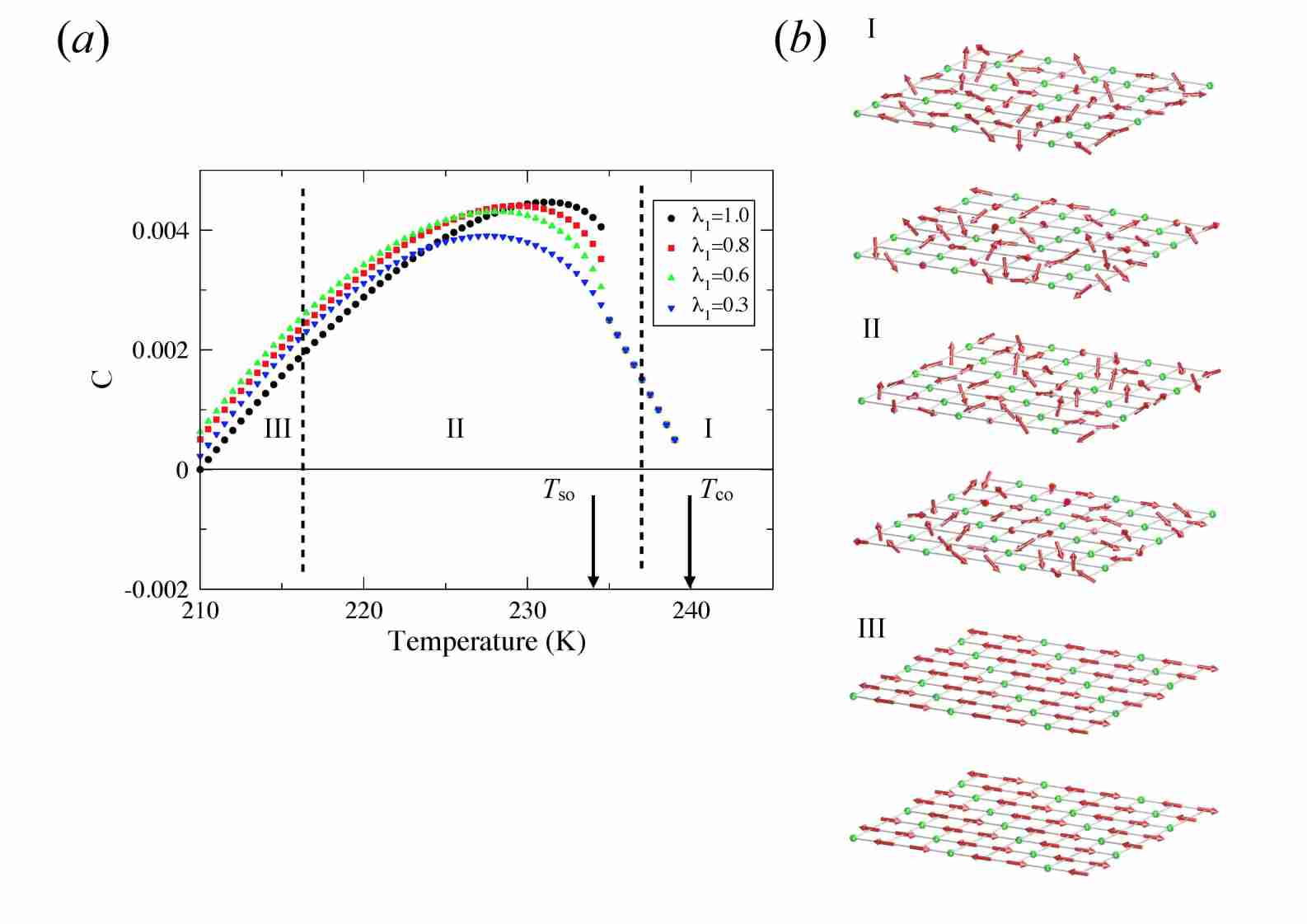}
\caption{Bilayer correlations from the Landau theory, and spin and charge 
configurations in different phases.  (a) Plot of C, the measure of the interlayer coupling strength defined in the text, as a function of temperature for several intra-layer charge-spin coupling constants. (b) Real-space configurations of the states in different temperature regimes. In the high temperature regime I, there exists no or very weak in-plane charge order and the correlation between layers is small. In regime II, charge stripe order develops while the spins remain disordered. The charge stripes between layers are out of phase to minimize the Coulomb repulsion between layers. At lower temperatures in regime III, in-plane spin stripes form and the interlayer charge correlation is suppressed.}
\label{fig3}
\end{figure*}
 
Giving $r_\rho$ and $r_s$ linear temperature dependence and taking $\lambda_1$ to be temperature independent,  $C$ will always, after an initial increase, decrease as temperature is lowered, regardless of the exact values of the parameters.    
As shown in Fig.~\ref{fig3}a, $C$ behaves similarly to the interlayer correlation length (Fig.~\ref{fig1}d) for temperatures not too far from the onset of the stripe order.

 \section{Discussion}
\vspace{-1em}
Intuitively,  this curious rise and fall of interlayer correlation  is a result of two competing interactions ~\cite{Rosenthal:2014}.  The interlayer charge-charge  coupling favors the stripes in different layers to be out of phase because the charge stripes repel each other, while the interlayer charge-spin coupling favors in-phase stripes because the formation of in-plane spin modulation causes the dissipation of kinetic energy of the electrons.  In-plane charge order appears first, resulting in the buildup of an out-of-phase interlayer charge correlation, but as the in-plane spin stripe order starts to develop, the interlayer charge order is suppressed.

\begin{figure}
\includegraphics[width=.5\textwidth]{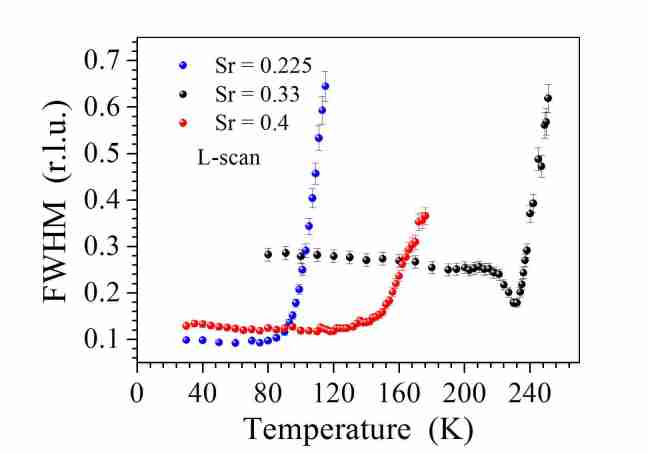}
\caption{Peak widths of charge stripes with different hole concentrations. The data were taken from single crystals of La$_{2-x}$Sr$_x$NiO$_4$ with different hole concentrations, i.e., $x = 0.225$, $x=0.33$, and $x=0.4$.  As can be seen, the inverse order-disorder transition occurs only for $x = 0.33$.}
\label{fig4}
\end{figure}

Figure~\ref{fig4} shows the measurements at  doping concentrations $x=0.225, 0.33$, and $0.4$; only at $x$ = ${1}/{3}$ does the inverse melting occur. This phenomenon adds to the list of anomalies for La$_{5/3}$Sr$_{1/3}$NiO$_4$ due to the commensurate pinning of the charges to the Ni lattice at $x$ = ${1}/{3}$~\cite{Ramirez:1996kq,Kajimoto:2001qf,Yoshizawa:2000xy,Ishizaka:2004tk,Abeykoon:2013ai}.   
For $x$ $\ne$ ${1}/{3}$, topological defects, such as dislocations and kinks, can easily proliferate to destabilise the in-plane charge stripe order~\cite{Li:2003qm, Lloyd-Huges:2008}. This also weakens the phase-dependent interlayer charge-spin couplings in Eq.~(\ref{eq:phase2}); as a result, there are no competing interlayer interactions to cause the inverse melting of the interlayer correlation. 

Our work points to the importance of the interlayer coupling in LSNO.
Interlayer Coulomb interaction  has been argued to be crucial in understanding an anomalous shrinking of the $c/a$ lattice parameter ratio that correlates with $T_{\rm CO}$ in Li$_{5/3}$Sr$_{1/3}$NiO$_4$~\cite{Abeykoon:2013ai}, as well as the existence of  fluctuating charge stripes that persist to  high temperatures~\cite{Du:2000,Abeykoon:2013ai}.  In particular, the inverse melting of the interlayer charge order observed in this work may provide a new direction to understand the dominance of the dynamical stripes in cuperates.  Further extension of the current work to study the dynamical interlayer correlations in  Li$_{5/3}$Sr$_{1/3}$NiO$_4$~\cite{Lee:2012ak,Abeykoon:2013ai} and its sister compound  Li$_{5/3}$Sr$_{1/3}$CoO$_4$~\cite{Boothroyd:2011, Lancaster:2014} may help to elucidate the physics of high-$T_c$ superconductors.

\begin{acknowledgments}
We acknowledge many stimulating discussions with Cheng-Hsuan Chen and Bruce Gaulin. We  are grateful to MOST in Taiwan for the financial support via NSC 99-2112-M-032-005-MY3 and NSC 102-2112-M-032-004-MY3 (CHD), NSC 102-2112-M-002 -003 -MY3 (YJK). 

\vspace{1em}
\textbf{Author Contributions:} PDH, YXS and CHD initiated the research. DP was in charge of the growth of single crystals. SHL and YCL performed the x-ray scattering works and characterizations of the crystals. DJH contributed to the resonant soft x-ray scattering. YJK and AFS carried out the theoretical calculations. YJK, AFS and CHD wrote the paper. CHD coordinated the project.  
\end{acknowledgments}

\appendix*

\begin{figure}
\includegraphics[width=.5\textwidth]{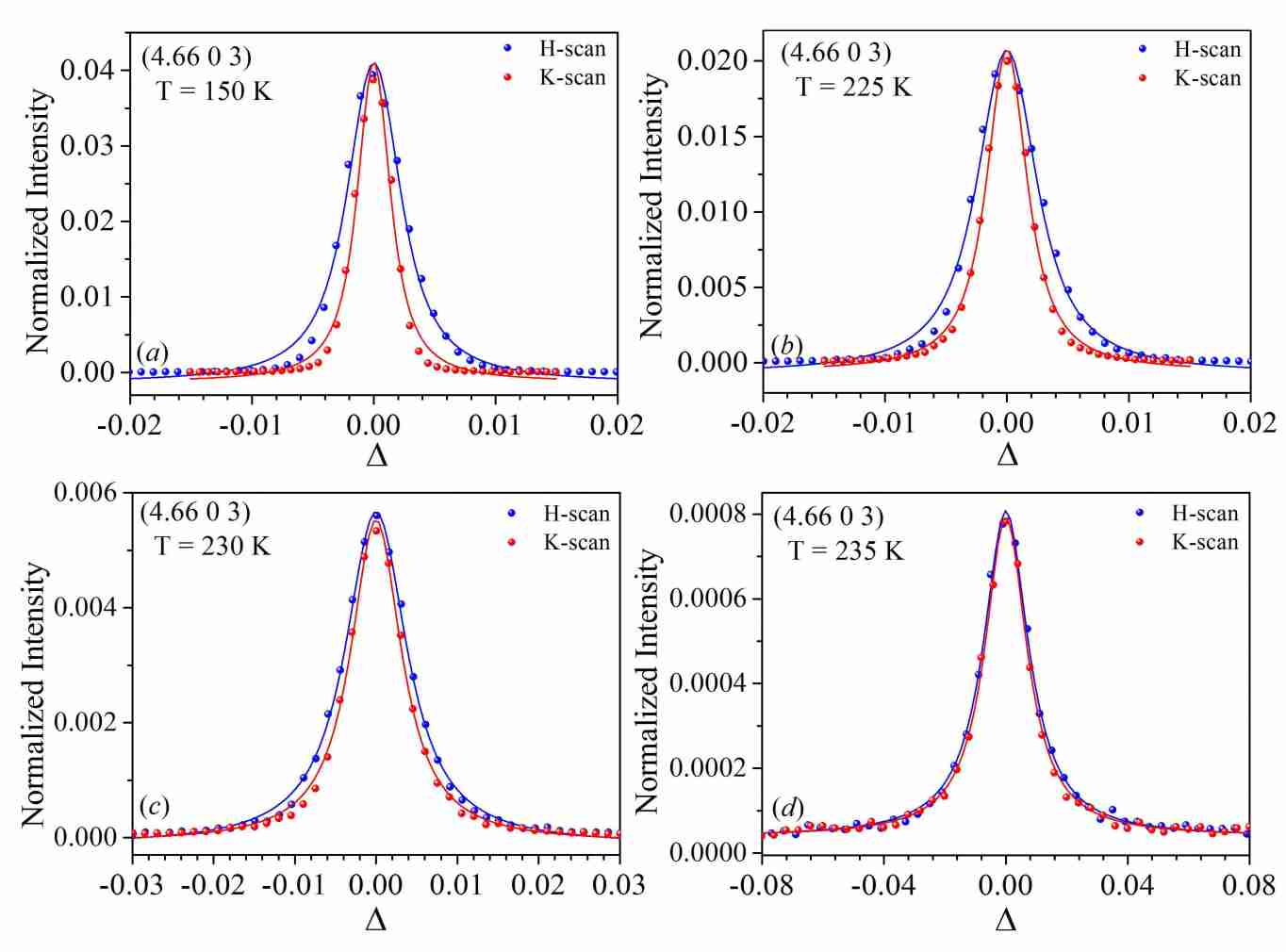}
\caption{Comparison of the peak profiles along the $H$- and $K$-direction at different temperatures. In order to compare the peak profiles along the $H$- and $K$-direction, the central positions of the charge stripe reflection $(4.66~ 0~ 3)$ are set to zero. (a) Below temperature $T\sim 218$ K, the ratio of peak widths along $H$- and $K$-directions is almost constant. (b) and (c) Upon warming, the ratio changes as a function of temperature, (d) approaching 1 as the temperature approaches $T_{CO}$ ($\sim$238 K).}
\label{figS1}
\end{figure}  

\begin{figure}
\includegraphics[width=.5\textwidth]{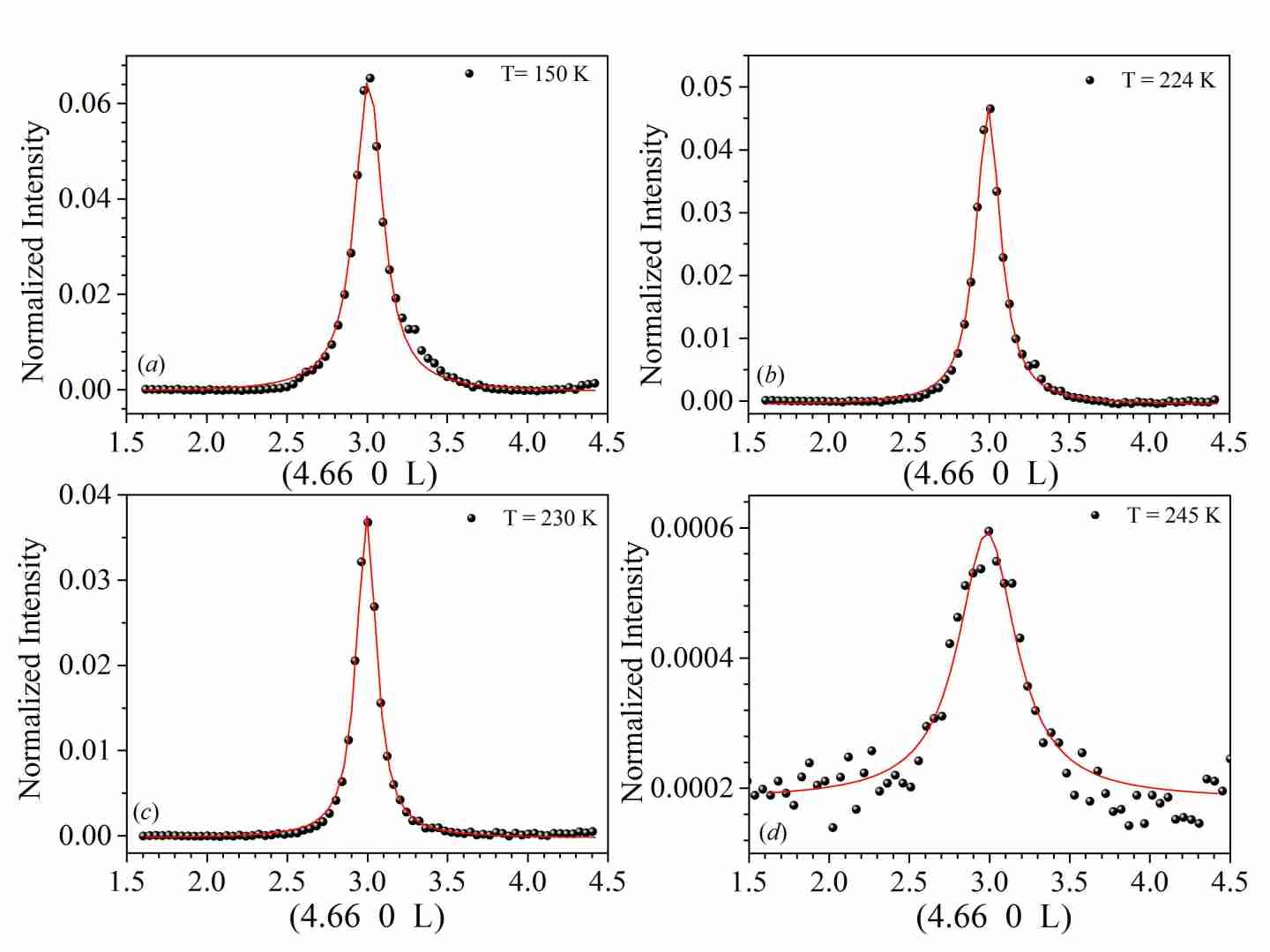}
\caption{Peak profiles of a charge stripe reflection along the $L$-direction at 
different temperatures. Scans through the $L$-direction ($c$-axis) of the charge stripe 
reflection $(4.66 ~0~ 3)$ at (a) 150 K, (b) 224 K , (c) 230 K, and (d) 245 K are shown. 
As temperature is lowered, the peak narrows and becomes sharpest at $\sim$230 K, but it 
then widens below 230 K, indicating an inverse melting transition.}
\label{figS2}
\end{figure}  

\begin{figure}
\includegraphics[width=.5\textwidth]{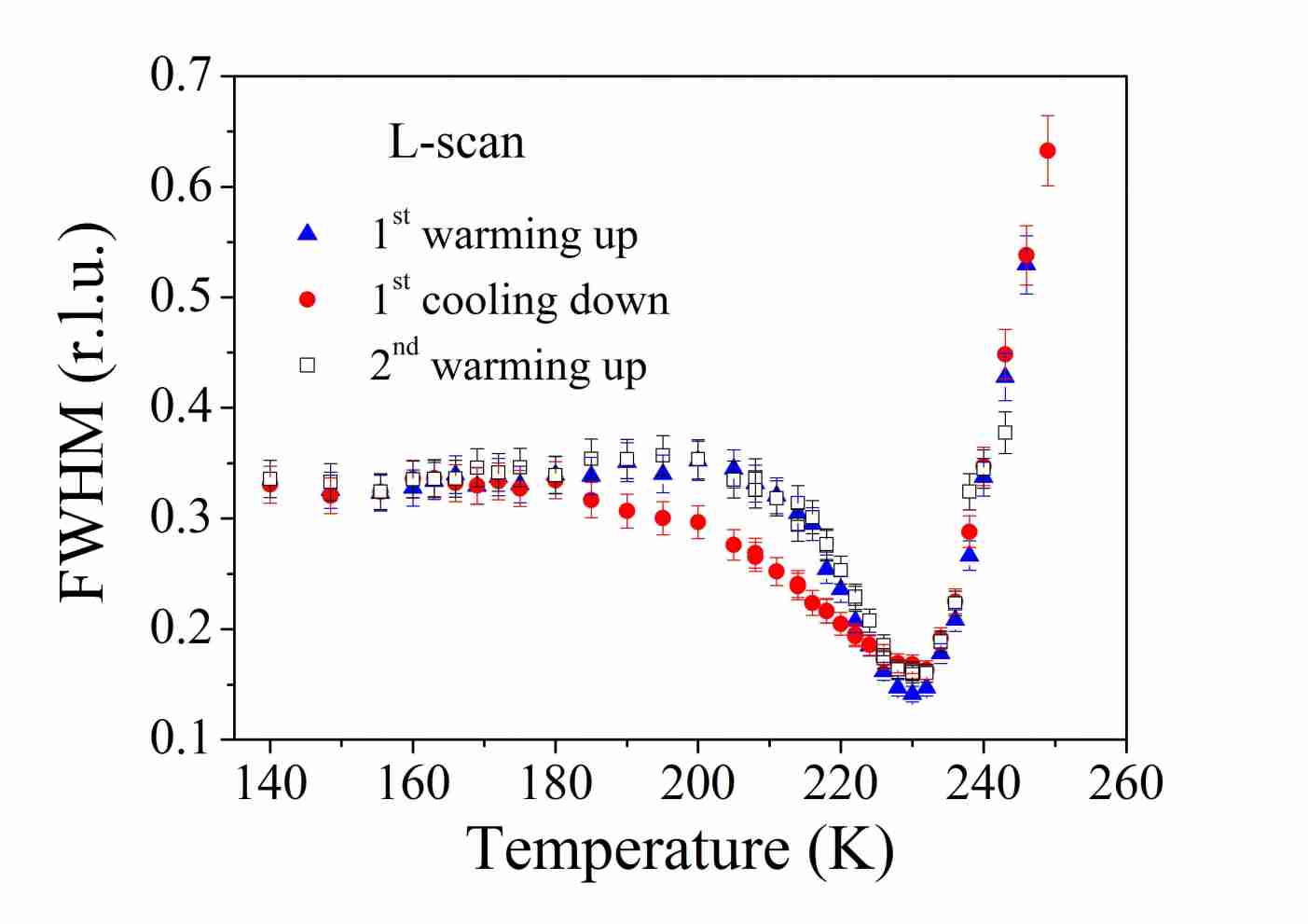}
\caption{ Thermal effects on the charge stripe modulations. Evolution of the peak 
width (FWHM) of the charge stripe reflection $(4.66 ~0~ 3)$ along the $L$-direction is 
shown as a function of temperature for the different thermal processes (see the 
explanation in the supplementary method description). As can be seen, there is some 
thermal hysteresis.}
\label{figS3}
\end{figure}  

\section{Methods} High quality single crystals of La$_{1-x}$Sr$_{x}$NiO$_4$, ${x}$= {0.225}, {1/3}, and {0.4}, were grown by the floating zone method. The crystals were characterized and orientated using conductivity measurements and an in-house x-ray diffractometer. The values of ${x}$ were further confirmed by checking the transition temperatures of charge modulation using synchrotron x-ray scattering.  Synchrotron x-ray scattering experiments were performed on the beamlines BL07 and SP12B1 of NSRRC (Taiwan) and SP8 (Japan). The sample was mounted on a closed-cycle cryostat  on a multi-circle diffractometer. A single crystal of LiF (0 0 1) was used as an analyzer to define the scattered x-rays from the sample. The experimental resolution function was determined to be $\epsilon_H^{-1}\sim0.0019 $ \AA$^{-1}$, $\epsilon_K^{-1}\sim0.001 $ \AA$^{-1}$, and $\epsilon_L^{-1}\sim0.015 $ \AA$^{-1}$ as measured on the Bragg peak (4 0 0) near the charge ordering peaks  at $T = 140$ K, with the sample mosaic width $\sim0.02^\circ$.
For the study of spin stripes, resonant soft X-ray scattering measurements were performed on the beamline BL05B3 of NSRRC. The measurements were performed to scan the spin stripe reflection (0.66 0 0) through the $L$ edge of Ni.

\vspace{1em}
\textbf{Charge and Magnetic Correlations.}  Synchrotron X-ray scattering experiments were carried out on the beamlines BL07 and SP12B1 of NSRRC, Taiwan. The sample was mounted on a closed-cycle cryostat mounted on a multi-circle diffractometer, which allows the scans to be performed along any of the reciprocal space crystallographic axes, $H (= 2\pi/a)$, $K (= 2\pi/b)$, and $L (= 2\pi/c)$. Throughout this study for La$_{5/3}$Sr$_{1/3}$NiO$_4$, a tetragonal unit cell with lattice parameters of $a = b = 5.4145$ \AA\ $=2\sqrt{2}d_{\rm Ni-O}$  and $c = 12.715$ \AA\  was used to index the reflections. There was no realignment of the crystal during the measurements because  La$_{5/3}$Sr$_{1/3}$NiO$_4$ does not display any structural phase transitions at low temperatures\cite{Abeykoon:2013ai}. The incident x-ray energy was set to 10 keV by a pair of high quality single crystals of Si(1 1 1), and a LiF crystal was used in an analyzer to define the scattered x-rays from the sample. The experimental resolution function was determined to be $\epsilon_H^{-1}\sim0.0019 $ \AA$^{-1}$, $\epsilon_K^{-1}\sim0.001 $ \AA$^{-1}$, and $\epsilon_L^{-1}\sim0.015 $ \AA$^{-1}$ as measured on the Bragg peak (4 0 0), which is near the charge ordering peaks measured at $T = 140$ K, and the sample mosaic width was found to be $\sim0.02^\circ$. The peak profiles of the Bragg reflection (4 0 0) were monitored throughout the measurements and showed no changes. The correlation lengths of the charge stripe reflections were extracted from their measured peak profiles convoluted with the resolution functions.  Measurements  were taken as a function of temperature through the Bragg peak and charge stripe satellites along the crystallographic axes of $H$, $K$, and $L$ in the reciprocal space. Figure~\ref{figS1} shows how the peak widths of the charge modulation along the $H$- and $K$-directions change as a function of temperature.  As can be seen, for temperatures above T$_{CO}$, the charge modulation is isotropic in the $a$$\times$$b$ plane, but as temperature is lowered, there is an anisotropic evolution of the correlation lengths.  Figure~\ref{figS2} displays the evolution of the peak profile of charge modulation along $c$-axis as a function of temperature.  As temperature is decreased, the peak narrows at first, indicating an increase in order along the $c$-axis, but then it widens again, indicating an inverse order-disorder transition.
  
For the study of spin stripes, in order to enhance the signals from the spin modulations, a resonant soft x-ray diffraction experiment was conducted on the beamline BL05B3 of NSRRC. The measurements were performed to scan the spin stripe reflection (0.66 0 0) through the $L$ edge of Ni. A large resonance from the spin reflection was observed at the $L_3$ edge of Ni with incident $\pi$-polarized x-rays. The data shown in Fig.~\ref{fig2}, marked as red dots, was collected as a function of temperature.

\vspace*{1em}
\textbf{Thermal Hysteresis.}  Since the charge stripes possess the characteristics of a liquid crystal, experiments were also conducted to study thermal effects on the charge modulation. As shown in Fig.~\ref{figS3}, charge stripes show a hysteresis behavior around the transition boundary between nematic and smectic phases under different thermal treatments. This is in accordance with previously described thermal phenomena of electron nematics\cite{Dahmen:2010}.  
The data shown in Fig.~\ref{figS3} were collected during three sequences of warming and cooling. The sample was first cooled down to $130$ K from room temperature in approximately 2 hours, and after the alignment at $130 K$, the data (as marked by blue triangles in Fig.~\ref{figS3}) were collected by increasing temperature and scanning the charge stripe reflection (4.66 0 3) along the $H$, $K$, and $L$ directions as a function of temperature until $T$= 250 K, where the reflection becomes very broad and weak.  The sample was then warmed up to 260 K and kept at that temperature for approximately half an hour, after which measurements (marked by red dots) were taken as the sample was cooled to $T$= 140 K.  A third round of measurements (marked by open squares) were taken as the sample was warmed up to 250 K once more.

\end{document}